# Structural and superconducting parameters of highly compressed sulfur

Evgeny F. Talantsev[1,2], Evgeniya G. Valova-Zaharevskaya[1],

[1]M. N. Miheev Institute of Metal Physics, Ural Branch, Russian Academy of Sciences, 18, S. Kovalevskoy St., Ekaterinburg 620108, Russia
[2]NANOTECH Centre, Ural Federal University, 19 Mira St., Ekaterinburg 620002, Russia

**Abstract**

Sulfur was the first nonmetal element which was transformed to a superconductor by applying megabar pressure. Recent pioneering experimental developments in measuring the superconducting energy gap $\Delta(T)$ in compressed sulfur using tunneling spectroscopy (Du et al., *Phys. Rev. Lett*. **133**, 036002 (2024)) initiated an interest in better understanding real atomic structure and superconducting properties of this element at high pressure. Here, we analyzed available experimental data on highly compressed sulfur, and, from the $\Delta(T)$ data reported by Du et al. (2024), we extracted the specific heat jump at the transition temperature of $\Delta C_{el}/\gamma T_c = 1.8$. We also developed a model to extract the Debye temperatures $\Theta_D$ for sulfur and $H_3S$ in two-phases sample from the temperature-dependent resistance $R(T)$. for better understanding of material structure, here we proposed to use a size-strain map for highly compressed samples, and we revealed this size-strain map for laser-heated sulfur in a diamond anvil cell with a mixture of sulfur and $H_3S$. Finally, we found that superconducting sulfur exhibits a moderate level of nonadiabaticity $0.04 \leq \Theta_D/T_F \leq 0.15$ (where $T_F$ is the Fermi temperature), which is similar to $MgB_2$, pnictides, cuprates, $La_4H_{23}$, $ThH_9$, $H_3S$, $LaBeH_8$, and $LaH_{10}$.

# Structural and superconducting parameters of highly compressed sulfur

## I. Introduction

Yakovlev *et al.*[1] were the first to transform insulating element into a superconductor by applying high pressure. This non-metallic element was sulfur. Since then, highly pressurized sulfur and other non-metallic elements, which transform into superconductors by applying high pressure, were studying for nearly five decades[2–17]. A revival of interest in superconducting sulfur followed the landmark discovery of near-room-temperature superconductivity in highly compressed $H_3S$[18], where elemental sulfur can be used as a starting material for the synthesis of the $H_3S$ phase [19–24]. The phase is in a focus of extended experimental and theoretical studies[22,25–46].

Very recent interest in highly compressed sulfur originates from several studies[7,12–15,23]. Here, in attempt to improve our understanding of the extremely diverse phase diagram and physical properties of highly compressed sulfur, we have performed further analysis of experimental data reported in four recent studies:

(1) Du *et al*[12] reported a ground-breaking experimental development to measure the superconducting energy gap, $\Delta(T)$, in a superconductor (e.g., sulfur) at megabar pressure using tunnelling spectroscopy technique;

(2) Wang *et al*[13] reported that at some pressure range, $P$, highly compressed sulfur has in-field temperature dependent resistance, $R(T,B,P)$, similar to those in underdoped ambient pressure cuprates;

(3) Osmond *et al*[23] reported experimental data on highly compressed two-phase sample $H_3S+S$ from which the properties of elemental sulfur have been deduced in the current study;

(4) Shi et al[14] further extended the phase diagram of sulfur by investigating the structure of sulfur after fast ramp compression.

## II. Data sources and theoretical models

Analysis was performed for datasets which are freely available online[12–14,23]. The experimental conditions under which the X-ray diffraction scans of highly compressed sulfur were obtained are given in the corresponding experimental parts of the original reports. The raw X-ray diffraction data were first analyzed using Dioptas[47] software. Descriptions of the developed model for temperature-dependent resistance and the well-known models used are given in the corresponding sections.

## III. Results and Discussion

### 3.1. Sulfur in a tunnel junction at $P = 160$ GPa.

Du et al.[12] measured temperature dependent gap amplitude, $\Delta(T)$, in sulfur compressed at $P = 160 \, GPa$. It should be noted that Gross et al.[48,49] proposed simple analytical equation for temperature dependent gap for the *s*-wave symmetry:

$$\Delta(T) = \Delta(0) \times tanh\left[\frac{\pi k_B T_c}{\Delta(0)} \times \sqrt{\eta \times \frac{\Delta C_{el}}{\gamma T_c} \times \left(\frac{T_c}{T} - 1\right)}\right], \tag{1}$$

where $k_B$ is the Boltzmann constant, $\eta = \frac{2}{3}$ for superconductors exhibited *s*-wave symmetry, $\frac{\Delta C_{el}}{\gamma T_c}$ is one of several universal unit-less ratios of the Bardeen-Cooper-Schrieffer[50] theory, which is the relative jump in the electronic heat at $T_c$, and $\gamma$ is the Sommerfeld parameter.

Based on this, we fitted the $\Delta(T)$ dataset[12] to Eq. 1 (where $\Delta(0)$, $T_c$, and $\frac{\Delta C_{el}}{\gamma T_c}$ are free fitting parameters) to deduce unknown fundamental ratio $\frac{\Delta C_{el}}{\gamma T_c}$ for highly compressed sulfur, in addition to $\Delta(0)$ and $T_c$. The fit is shown in Fig. 1 and deduced ratios are $\frac{\Delta C_{el}}{\gamma T_c} = 1.8 \pm 0.1$ and $\frac{2\Delta(0)}{k_B T_c} = 3.89 \pm 0.06$. It is evident that the obtained value $\frac{2\Delta(0)}{k_B T_c}$ is within the uncertainty intervals for the value $\frac{2\Delta(0)}{k_B T_c} = 3.8$ reported by Du et al.[12].

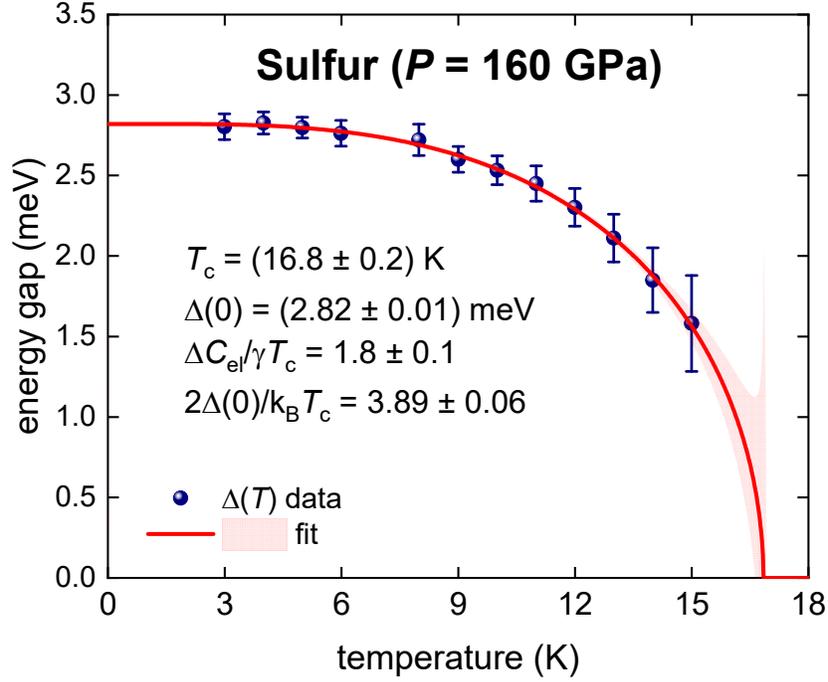

**Figure 1.** Superconducting gap values, $\Delta(T)$, in compressed sulfur ($P$ = 160 GPa) measured by Du et al.[12] and data fit to Gross et al.[48,49] model. Deduced values are shown. 95% confidence intervals are indicated by the pink shaded area. Fit quality R-Squared (COD) = 0.9948.

To demonstrate that deduced $\frac{\Delta C_{el}}{\gamma T_c}$ and $\frac{2\Delta(0)}{k_B T_c}$ for highly compressed sulfur (Fig. 1) are well aligned with general trend established for 34 classical ambient-pressure electron-phonon superconductors by Carbotte[51], in Fig. 2 we showed the datapoint for compressed sulfur together with full dataset reported by Carbotte[51].

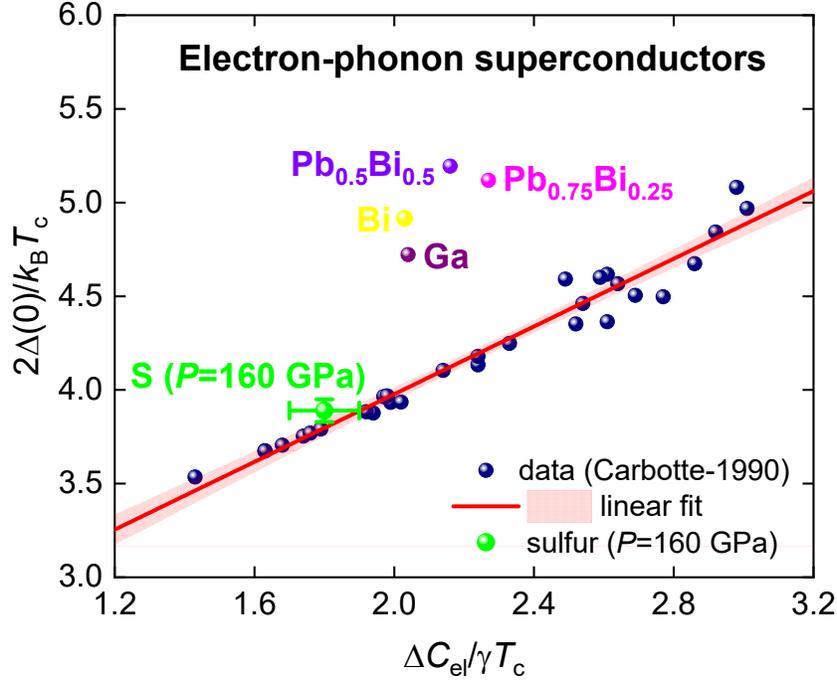

**Figure 2.** $\frac{2\Delta(0)}{k_B T_c}$ vs $\frac{\Delta C_{el}}{\gamma T_c}$ datapoint for highly compressed sulfur (*P*=160 GPa) derived herein and dataset for 34 ambient pressure classical electron-phonon superconductors reported by Carbotte[51]. Data for Ga, Bi, Pb$_{0.5}$Bi$_{0.5}$, Pb$_{0.75}$Bi$_{0.75}$ and S(*P*=160 GPa) are not included in the linear fit. 95% confidence intervals are indicated by the pink shaded area.

## 3.2. Hysteretic $R(T,B,P)$ curves of sulfur at *P* = 93-196 GPa

Wang *et al*[13] made publicly available several raw $R(T \leq 280\ K, B \leq 9\ T, 93\ GPa \leq P \leq 196\ GPa)$ datasets for sulfur which we examined in details herein. In their paper, Wang *et al*[13] reported on the negative slope of the resistance in sulfur-phase-V at applied field $0\ T \leq B \leq 1\ T$ and temperature range of $T = 2 - 30\ K$.

Closer examination of raw $R(T,B,P)$ datasets[13], however, revealed a more complicated picture. For instance, in Fig. 3 we showed full $R(T, 0\ T \leq B \leq 1.0\ T, P = 163\ GPa)$ dataset, which was only partially shown by Wang *et al*[13] in their Fig. 2,a[13].

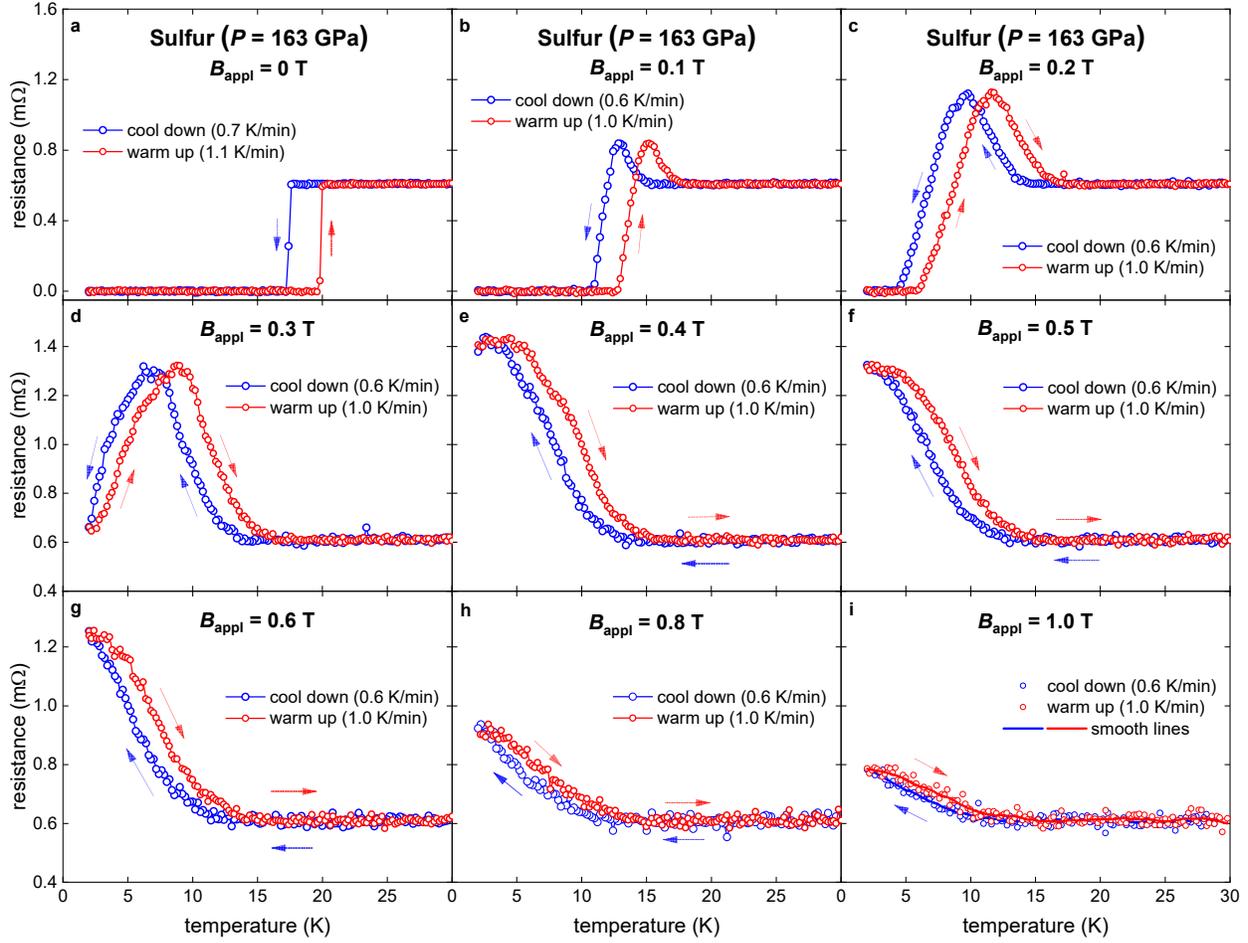

**Figure 3.** $R(T, B, P = 163\ GPa)$ in compressed sulfur measured by Wang *et al*[13] and partially reported in Fig. 2,a[13]. All experiments started at $T = 30\ K$ at the settled magnetic field $B_{appl}$, and then the sample was cooled down to $T = 2\ K$ and warmed up with indicated rates.

It is evident from Fig. 3, that the $R(T, B, P)$ curves demonstrate prominent hysteresis.

We can propose/discuss several explanations for the observed hysteresis (Fig. 3). First of all, we need to stress that elemental sulfur exhibits more than 30 allotropic phases at normal conditions[52] and, thus, any observed hysteresis for some physical property can be associated with one of many possible phase transitions. The second, as it was mentioned by Ma *et al*.[53] the pressure in DAC has trend to increase in the thermal cycling. Based on this, the *R(T,B,P)* measurements as a rule are performing on the warming semi-cycle[53]. The third reason for the observed hysteresis in Fig. 3 is that the cooling-warming rates (~ 1 K/min) used by Wang *et al*[13] were too fast that the sample and the DAC can reach thermal equilibrium with the PPMS chamber. This hypothesis can get further support by considering full raw $R(T, B = 0, P = 93\ GPa)$ dataset[13] (Fig. 4), which was partially showed by Wang et al[13] in their Fig. 1[13]. One can see the large difference between the $R(T, B = 0, P = 93\ GPa)$ measured at cooling rate of 0.7 K/min and warming rate of 2.0 K/min. However, we should pointed out again, that the issue with different pressure at the cooling and warming stages[53] in DAC can be the reason for the observed hysteresis in Fig. 4.

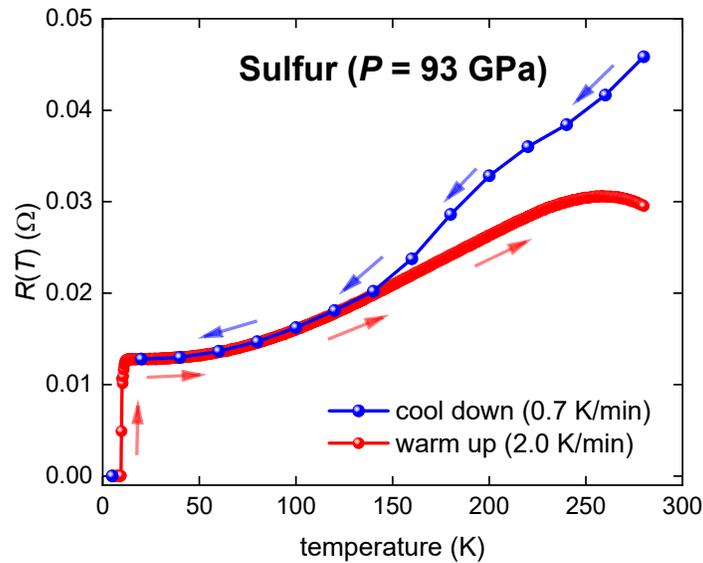

**Figure 4.** $R(T, B, P = 93\ GPa)$ in compressed sulfur measured by Wang et al[13] and partially reported in Fig. 1,a[13]. The sample at $B_{appl} = 0$ was cooled down from $T = 280\ K$ to $T = 2\ K$ with rates of 0.7 K/min, and then warmed up to $T = 280\ K$ with the rate 2.0 K/min.

The fifth probable explanation is that, at this temperature-pressure range and used cooling-warming rates, the compressed sulfur exhibits two-phase state, where one phase can be one-dimensional charge-density wave (CDW) phase with transition at $T \sim 150$ K (this is just an estimate based on Fig. 4). CDW transition in compressed sulfur was proposed by Degtyareva et al.[10]. The CDW can affect the $R(T, B, P)$ in a hysteretic manner, similar to hysteretic $R(T)$ curves[54–56] reported for the IrTe2, where stripe-charge ordering and phase structural transitions occur at $T < 300\ K$ play a dominant role in the emergence of the superconductivity[57] in IrTe$_2$. Based on this, we propose that the cooling-warming conditions in high-pressure studies need to be reported, and the rate for the cooling-warming cycle should be lower than 2.0 K/min.

To some extent, all $R(T, P)$l curves measured by Wang *et al*[13] in highly compressed sulfur are hysteretic (which can be seen in raw data[13]). Considering a fact that the same team performed high-pressure resistance measurements in Ref.[58], more experimental data is required to results reported in Ref.[58], because the $R(T, P)$ curves in Ref.[58] have been reported at the cooling semi-cycle only, and there is no data for the warming part of all cycles.

Below we analyzed the $R(T, B = 0, P)$ datasets[13] measured at the warming stage, because the $R(T, B = 0, P)$ datasets[13] at the cooling stage do not have enough raw data for the analysis, and pressure stabilization in DAC at this semicycle[53].

### 3.3. Pressure dependence of the transition temperature, Debye temperature and the electron-phonon coupling constant

Bardeen-Cooper-Schrieffer (BCS) theory[50] predicts that pressure should suppress the $T_c$ in simple metals (with *s*- and *p*-electrons), because the BCS postulate that the superconducting phenomenon is based on the electron-phonon pairing and based on this[11]:

$$T_c(P) \propto \sqrt{\frac{k(P)}{M}} \times e^{-\frac{k(P)}{\eta(P)}}, \qquad (2)$$

where $M$ is the ion mass, $k(P)$ is pressure dependent lattice spring constant, and $\eta(P)$ is pressure dependent Hopfield parameter[59] (which is proportional to the electronic density of states and the mean-square electron-phonon matrix element). Due to slow pressure dependence of the $\eta(P)$ in comparison with the $k(P)$ in simple metals, applied pressures cause the suppression of the $T_c(P)$, because of fast suppression of the exponential term in Eq. 2.

The $T_c(P)$ dependences for elements which were metallized under applied pressure (like, sulfur[2], oxygen[60], and others[5,61]) or compounds[18,62–64] synthesized in the DAC do not cover by universal theory. In each of these cases, first-principles calculations[30,65–67] are used to predict the $T_c(P)$ relationship that can be compared with experiment.

Here we deduced pressure dependences for the Debye temperature, $\Theta_D(P)$, the electron-phonon coupling constant, $\lambda_{e-ph}(P)$, and the transition temperature, $T_c(P)$, in compressed sulfur. For the latter, we utilized strict resistance criterion[68–70]:

$$\frac{R(T=T_{c,0.01})}{R_{norm}} = 0.01, \qquad (3)$$

where $R_{norm}$ is the resistance at temperature, where the $R(T)$ starts to drop, and the temperature at this drop is $T_c^{onset}$.

In Figure 5 we showed the $R(T, B = 0, P)$ datasets reported by Wang et al[13] together with fits to the saturated resistance model[71,72], which is based on Bloch- Grüneisen[73,74] equation, and where the Debye temperature, $\Theta_D$, is one of free-fitting parameters[73–75]:

$$R(T) = \cfrac{1}{\cfrac{\theta(T_c^{onset}-T)}{\left(I_0\left(F\times\left(1-\frac{T}{T_c^{onset}}\right)^{3/2}\right)\right)^2} R_{norm}} + \theta(T-T_c^{onset})\times\left(\cfrac{1}{R_{sat}} + \cfrac{1}{R_{norm}+A\times\left(\left(\frac{T}{\Theta_D}\right)^N\times\int_0^{\frac{\Theta_D}{T}}\frac{x^N}{(e^x-1)(1-e^{-x})}dx - \left(\frac{T_c^{onset}}{\Theta_D}\right)^N\times\int_0^{\frac{\Theta_D}{T_c^{onset}}}\frac{x^N}{(e^x-1)(1-e^{-x})}dx\right)}\right) \qquad (4)$$

where $\theta(x)$ is the Heaviside step function, $N$ is the power-law exponent associated with the dominant type of the charge carrier interaction[75–81] in materials with metallic resistance (where $N = 5$ is associated with the electron-phonon interaction), $I_0(x)$ is the zero-order modified Bessel function of the first kind, and $T_c^{onset}$, $F$, $\Theta_D$, $R_{norm}$, $A$, $R_{sat}$ are free-fitting parameters.

The term of $\cfrac{R_{norm}}{\left(I_0\left(F\times\left(1-\frac{T}{T_c^{onset}}\right)^{3/2}\right)\right)^2}$ was proposed by Tinkham[82] (and further modified in Ref.[83]), and the term of $\left(\frac{1}{R_{sat}} + \frac{1}{R_{norm}+A\times\left(\frac{T}{\Theta_D}\right)^5\times\int_0^{\frac{\Theta_D}{T}}\frac{x^5}{(e^x-1)(1-e^{-x})}dx}\right)$ was proposed by Wiesmann et al.[72]. Further details for the fitting function (Eq. 4) can be found elsewhere[68,72,81–86], including our previous studies[80,83,87] where we analyzed data on highly compressed sulfur S-III phase reported in Refs.[2,61]. In Fig. 5 we used the phase identification (i.e., S-III, S-IV, and S-V) based on report by Degtyareva et al.[8], and other relevant studies[6,8–10,13,16].

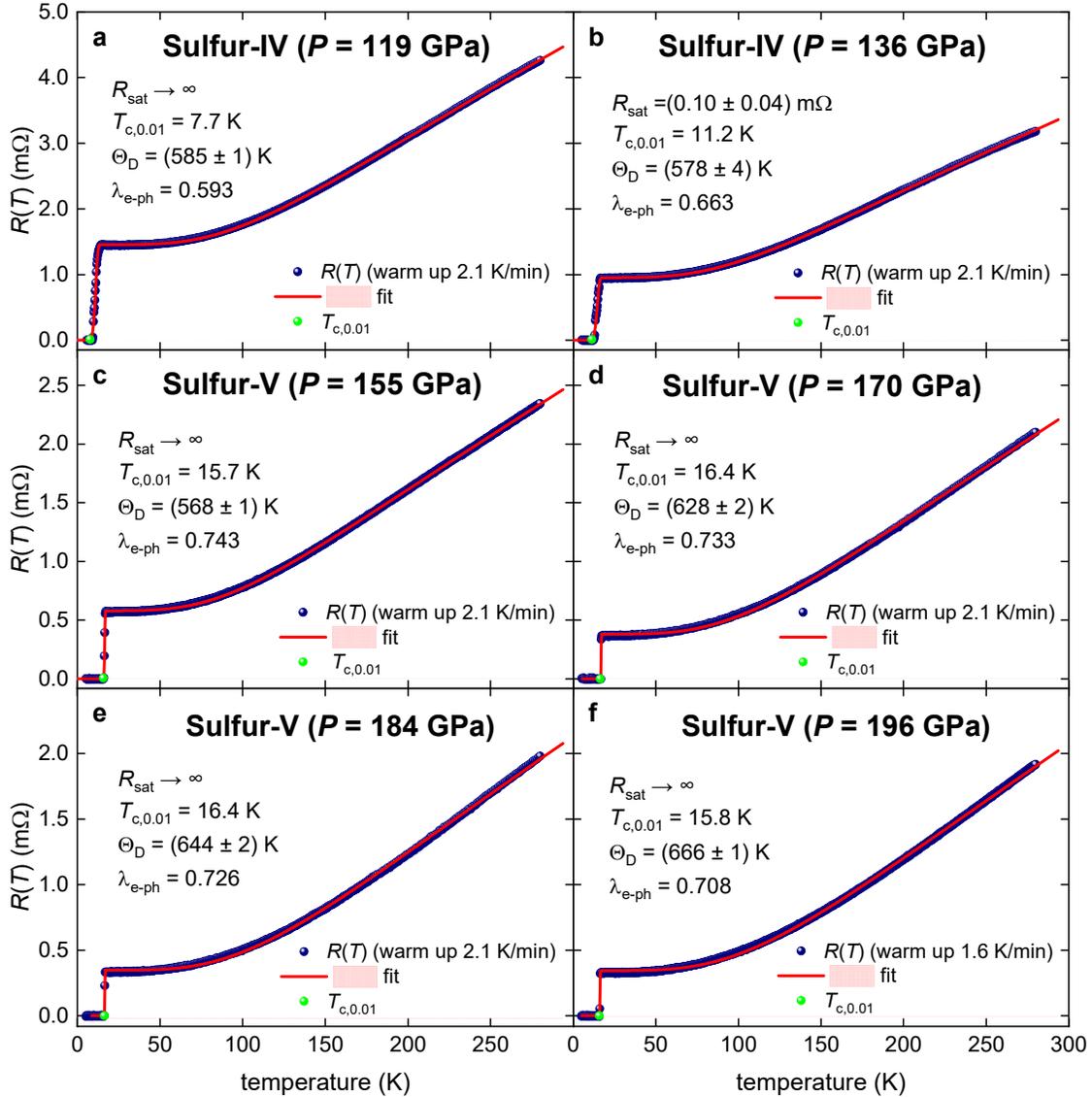

**Figure 5.** $R(T, B = 0, P)$ in compressed sulfur measured by Wang et al[13] and data fits to Eq. 4, where $N = 5$ is fixed parameter. Green balls indicate $T_{c,0.01}$. The goodness of fit: (a) 0.9997; (b) 0.9998; (c) 0.99992; (d) 0.9996; (e) 0.9995; (f) 0.9996. The thickness of 95% confidence bands (pink shadow areas) is narrower than the width of the fitting lines.

Deduced $\Theta_D$ and $T_{c,0.01}$ were used to determine the electron-phonon coupling constant $\lambda_{e-ph}$, which is the root of the system of equations[68,88,89]:

$$\begin{cases} T_{c,0.01} = \frac{\Theta_D}{1.45} \times e^{-\left(\frac{1.04(1+\lambda_{e-ph})}{\lambda_{e-ph}-\mu^* \times (1+0.62\times\lambda_{e-ph})}\right)} \times f_1 \times f_2^* \\ f_1 = \left(1 + \left(\frac{\lambda_{e-ph}}{2.46\times(1+3.8\times\mu^*)}\right)^{3/2}\right)^{1/3} \\ f_2^* = 1 + (0.0241 - 0.0735 \times \mu^*) \times \lambda_{e-ph}^2 \end{cases} \quad (5)$$

where $\mu^*$ is the Coulomb pseudopotential parameter (ranging from $\mu^* = 0.10 - 0.16$[66,90]), for which the mean value of $\mu^* = 0.13$ will be used for compressed sulfur below.

Derived $\Theta_D(P)$, $T_{c,0.01}(P)$, and $\lambda_{e-ph}(P)$ are shown in Fig. 6, where we also showed deduced values for S-III phase from our previous studies[83,87]. Derived $\lambda_{e-ph}(P \geq 119\ GPa)$ (Figs. 5,6) are in a very good agreement with the value computed by first principles calculations[7,91].

There is an immediate conclusion that the phase boundary between phase S-III and S-IV is the range of $110\ GPa < P < 119\ GPa$. This can be seen in the $\Theta_D(P)$ and $\lambda_{e-ph}(P)$ dependences, where a remarkably huge break in $\Theta_D(P)$ from $\Theta_D(P = 110\ GPa) \cong 200\ K$ and $\Theta_D(P = 119\ GPa) \cong 600\ K$ has been revealed.

We determined this phase boundary at $110\ GPa < P < 119\ GPa$, which is significantly different from the reported boundary at $P \cong 85\ GPa$ in Refs.[8,13]. However, our approach to determine the phase boundary is based on the derivation of the Debye temperature (and the electron-phonon coupling constant), which have a huge (large) change in the pressure range of $110\ GPa < P < 119\ GPa$, while the standard approach is based on the change in the slope of lattice parameters variation vs applied pressure and where the change is small[8,13].

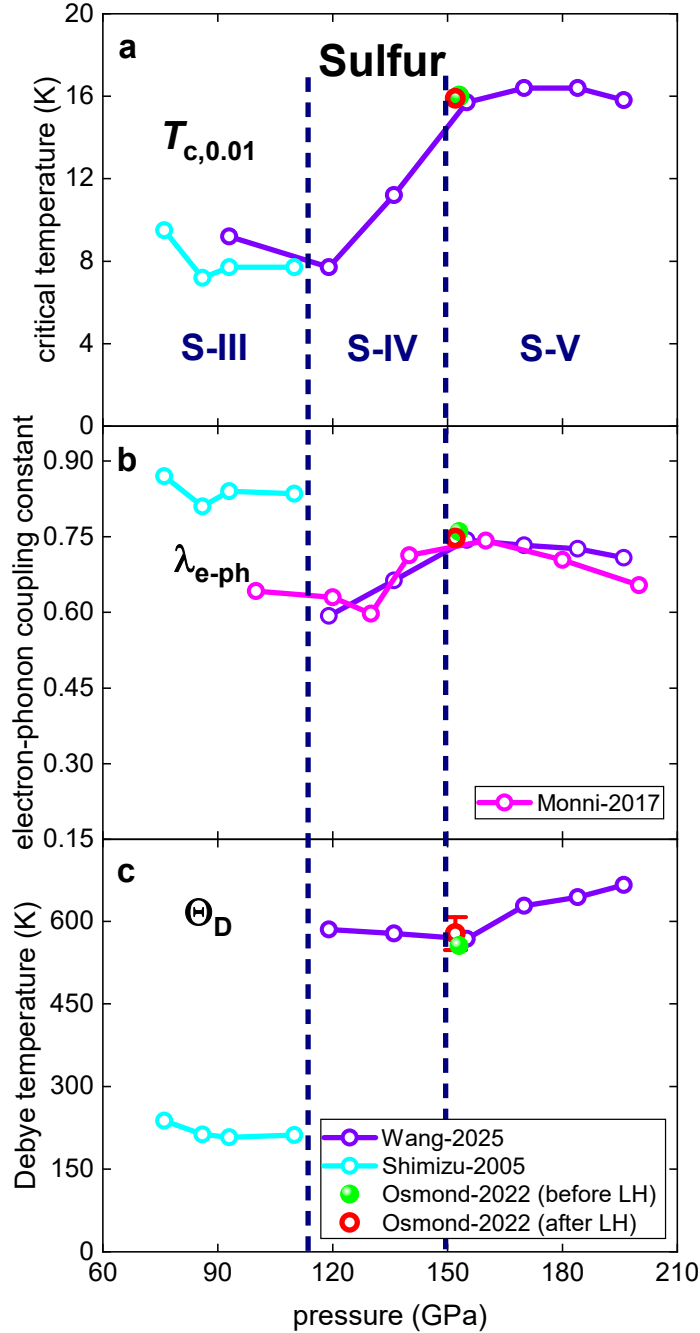

**Figure 6.** Derived (a) $T_{c,0.01}(P)$, (b) $\lambda_{e-ph}(P)$, and (c) $\Theta_D(P)$ for highly compressed sulfur. Magenta – $\lambda_{e-ph}(P)$ computed by the first-principles calculations by Monni et al.[91]; cyan – values deduced in Refs.[83,87] from experimental $R(T,P)$ data reported by Shimizu et al.[5]; violet – values deduced herein from experimental $R(T,P)$ data reported by Wang et al.[13]; green - value deduced herein from experimental $R(T, P = 153\ GPa)$ data reported by Osmond et al.[23] for pure sulfur before the laser heat (LH); red - value deduced herein from experimental $R(T, P = 153\ GPa)$ data reported by Osmond et al.[23] for mixture of $BH_3NH_3$ and sulfur after LH. Phase boundaries are based on Refs.[5,6,8–10,13,16,23].

### 3.4. The upper critical field and the Fermi temperature

Du et al.[12] reported the upper critical field data, $B_{c2}(T, P = 160\ GPa)$, which was deduced from measured $\Delta(T, B, P = 160\ GPa)$ dependence, as the field at which the superconducting gap closes. This $B_{c2}(T, P = 160\ GPa)$ dataset is shown in Fig. 7,a together with the fit to the $B_{c2}(T)$ analytical model recently proposed by Prozorov and Kogan[92]:

$$B_{c2}(T) = \frac{\phi_0}{2\pi\xi^2(0)} \times \left(\frac{1-\left(\frac{T}{T_c}\right)^2}{1+0.42\times\left(\frac{T}{T_c}\right)^{1.47}}\right), \quad (6)$$

where $\phi_0$ is the superconducting magnetic flux quantum, and $\xi(0)$ is the ground state coherence length.

In Fig. 7,b we showed the $B_{c2}(T, P = 163\ GPa)$ dataset which was deduced from the $R(T, B, P = 163\ GPa)$ reported by Wang et al[13] (in their Fig. 2,b[13]) by applying the criterion described by Eq. 3. The fit of the $B_{c2}(T, P = 163\ GPa)$ to the Eq. 7 is shown in Fig. 7,b.

Deduced $\xi(0)$ values were substituted to the following equation to calculate the Fermi temperature, $T_F$[41,93]:

$$T_F = \frac{\pi^2 m_e}{8 \cdot k_B} \times (1 + \lambda_{e-ph}) \times \xi^2(0) \times \left(\frac{\alpha k_B T_c}{\hbar}\right)^2, \quad (7)$$

where $m_e$ is bare electron mass, $\hbar$ is reduced Planck constant, $\alpha \equiv \frac{2\Delta(0)}{k_B T_c}$ is the gap-to-transition temperature ratio. Other parameters in Eq. 7 were taken from Figs. 1,6. For $B_{c2}(T, P = 163\ GPa)$ data showed in Fig. 7,b, the gap-to-transition temperature ratio $\frac{2\Delta(0)}{k_B T_c}$ was calculated by linear empirical relation proposed in Ref.[94]:

$$\alpha \equiv \frac{2\Delta(0)}{k_B \cdot T_c} = 3.26 + 0.74 \times \lambda_{e-ph}, \quad (8)$$

Calculated $T_F$ values are shown in both panels of Fig. 7.

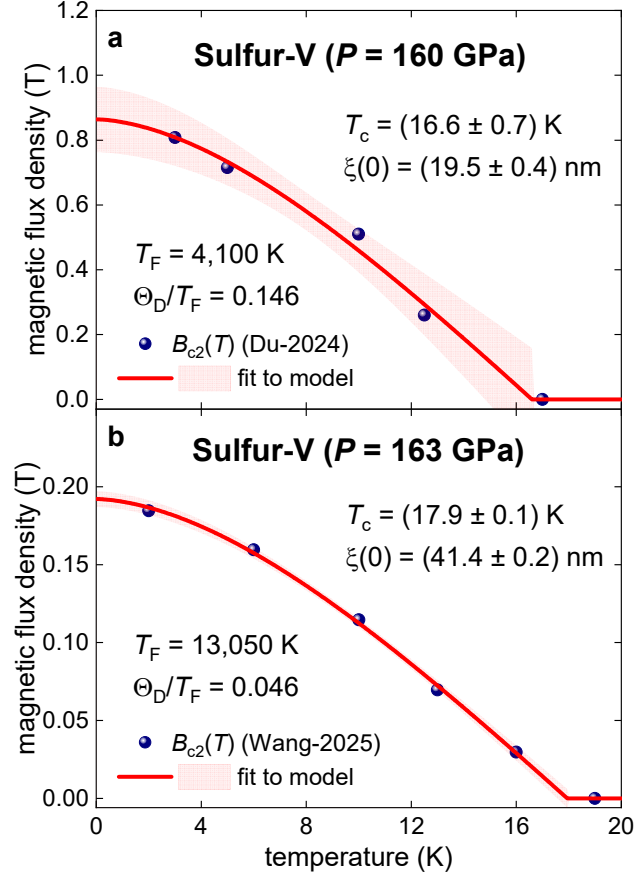

**Figure 7.** $B_{c2}(T)$ data and fits to to Eq. 6[92]. (a) $B_{c2}(T)$ reported by Du *et al*[12]; and (b) deduced from $R(T, B, P = 163\ GPa)$ data reported by Wang *et al*[13] by applying criterion of $\frac{R(T=T_{C,0.01})}{R_{norm}} = 0.01$. Fits quality (a) $R$ = 0.991; and (b) $R$ = 0.9992. Derived parameters are shown. The value of $\Theta_D = 600\ K$ is assumed.

Deduced Fermi temperature $T_F$ (Fig. 7) was used to locate sulfur-V phase in the Uemura plot[95,96] (Fig. 8). In the result, it was found that the sulfur phase S-V locates in a close proximity to the sulfur phase S-III for which we deduced $T_F$ in previous study[87]. It can be seen (Fig. 8) that both phases of sulfur are located at the intermediate band between unconventional and BCS bands in the Uemura plot.

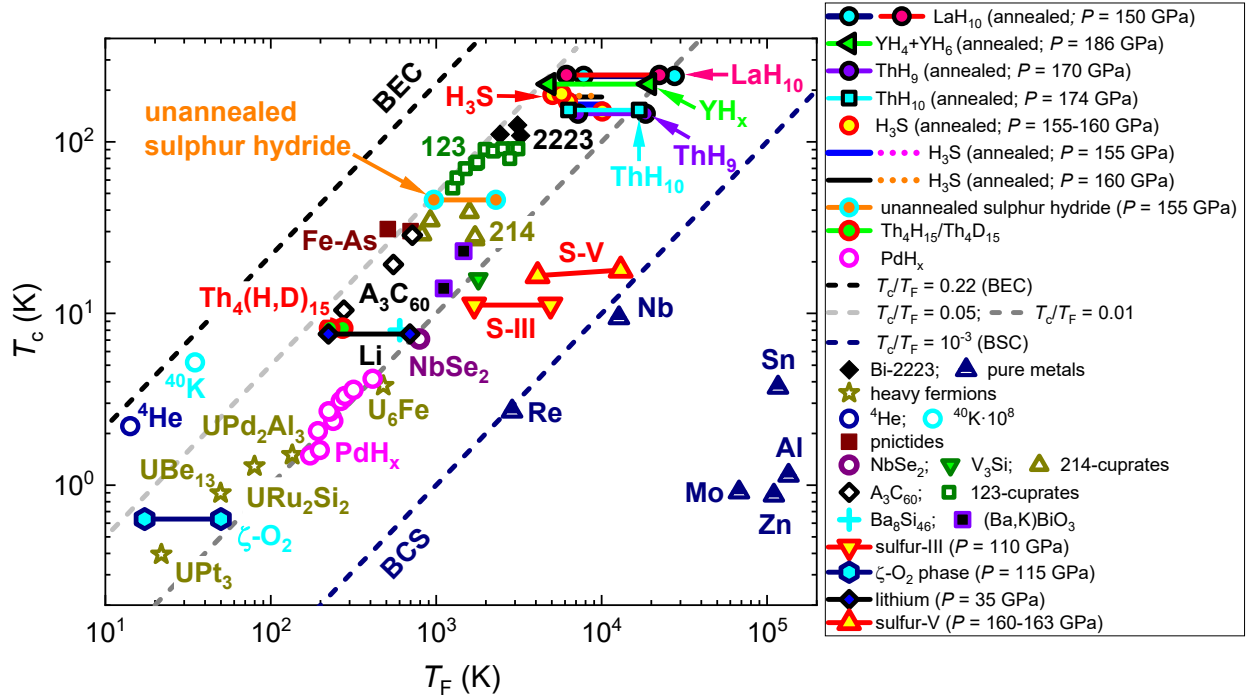

**Figure 8.** Uemura plot ($T_c$ vs. $T_F$) where superconducting phases of sulfur (S-III and S-V) are shown together with main superconducting families. References to the original data are in Refs.[85,93,97,98].

We also located S-III and S-V phases in the $\frac{\Theta_D}{T_F}$ vs $T_c$ plot (Fig. 9). This type of plot was proposed in Ref.[97] and it shows that all high-$T_c$ and near-room-temperature superconductors exhibit[85,97,98] the $\frac{\Theta_D}{T_F}$ ratio in the range of $0.04 \leq \frac{\Theta_D}{T_F} \leq 0.4$. Both sulfur phases studied herein exhibit the ratio of $\frac{\Theta_D}{T_F}$ in this range (Figs. 7,9).

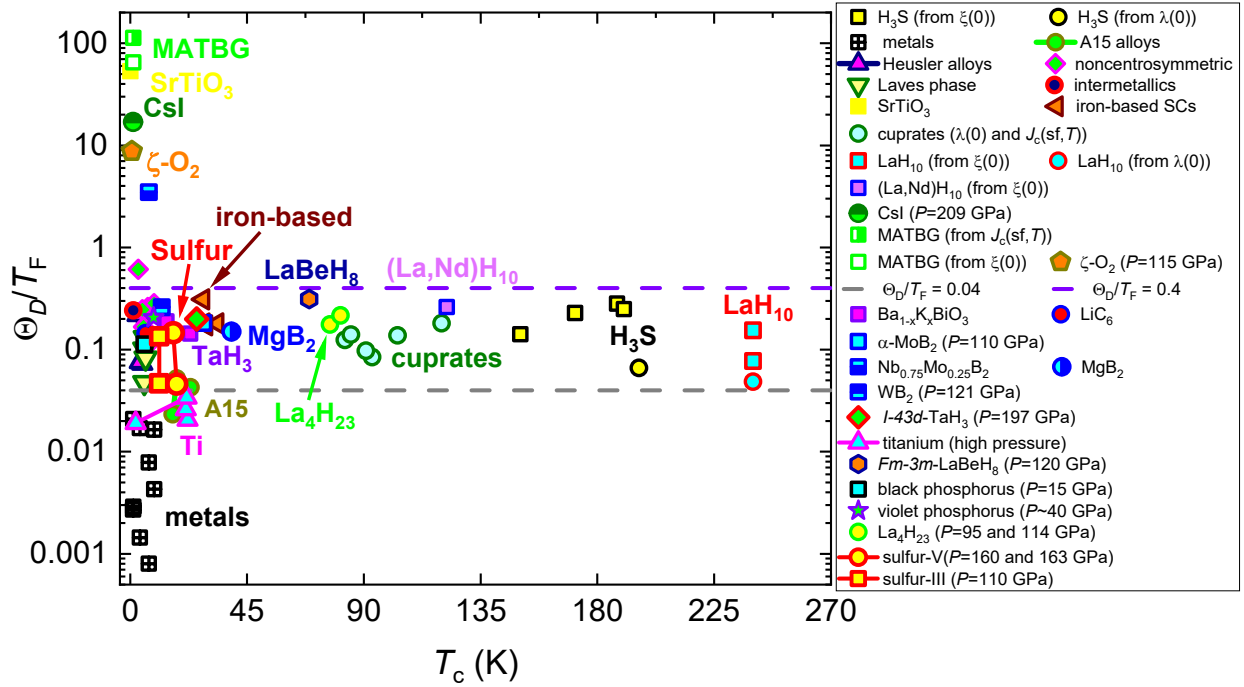

**Figure 9.** Plot of $\frac{\Theta_D}{T_F}$ vs $T_c$ for main families of superconductors, and where the superconducting phases of sulfur (S-III and S-V) are shown. References to data can be found in Refs.[41,85,93,97,98].

### 3.5. Highly compressed unreacted sulfur in the H₃S+S mixture before laser heating

Considering that sulfur is a starting chemical element for the synthesis of the $H_3S$ and $D_3S$ superconductors[19,20,22,23,31,32,45], it is important to analyse available experimental data on unreacted elemental sulfur in the mixture of $H_3S$, $H_3NH_3B$, and S in the DAC874 where the $H_3S$ phase is synthesized.

As the first step of this analysis, in Fig. 10 we fitted to Eq. 4 the $R(T, P = 153\ GPa)$ data measured in compressed sulfur before the laser heating (study by Osmond $et\ al$[23]). Deduced $T_{c,0.01}(P)$, $\lambda_{e-ph}(P)$, and $\Theta_D(P)$ from Fig. 10,a are shown in Fig. 6, where one can see excellent agreement between deduced values from raw data reported by two different research groups[13,23]. It should be also noted that the absolute values of the $R(T, P = 153\ GPa)$[23] and $R(T, P = 155\ GPa)$[13] are differed by a factor of ~ 70. This shows the robustness of our analysis (Eqs. 4,5 and Ref.[68]).

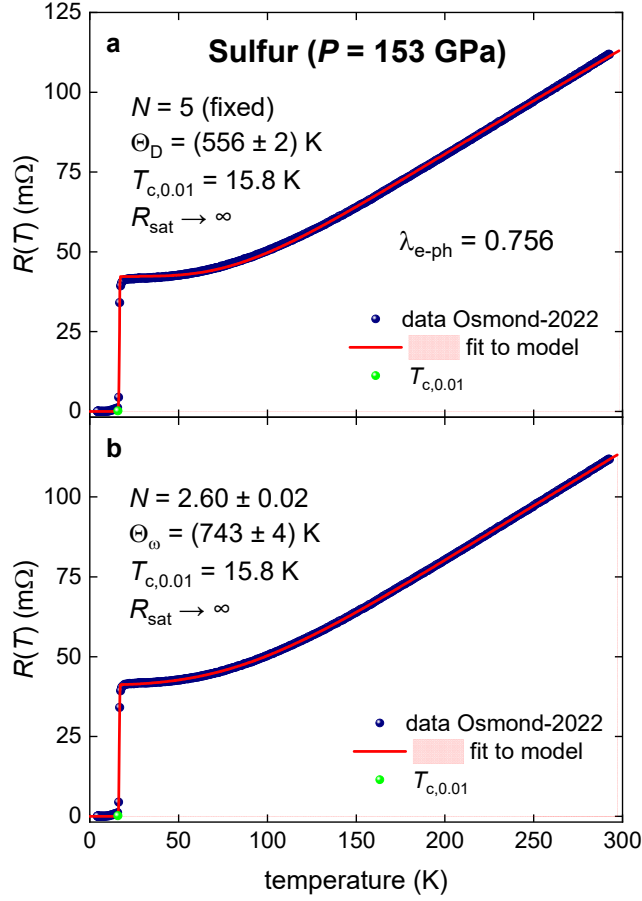

**Figure 10.** $R(T, B = 0, P = 153\ GPa)$ in compressed sulfur measured by Osmond *et al*[13] and data fits to Eq. 4, where (**a**) $N = 5\ (fixed)$ and (**b**) $N = 2.60 \pm 0.02$ is free-fitting parameter. Deduced parameters are shown. Green balls indicate $T_{c,0.01}$. The goodness of fits: (a) 0.9997 and (b) 0.99997. The thickness of 95% confidence bands (pink shadow areas) is narrower than the width of the fitting lines.

Osmond *et al*[23] also reported that they fitted normal state resistance data $R(T, P = 153\ GPa)$ to the Bloch- Grüneisen equation, where the power-law exponent $N$ was fixed to 3:

$$R(T) = R_{norm} + A \times \left(\frac{T}{\Theta_\omega}\right)^3 \times \int_0^{\frac{\Theta_\omega}{T}} \frac{x^3}{(e^x-1)(1-e^{-x})}\,dx. \qquad (9)$$

where $\Theta_\omega$ is the characteristic temperature associated with the power-law exponent $N = 3$, which differs from the Debye temperature $\Theta_D$ which is associated with the electron-phonon scattering and $N = 5$. In the result, Osmond *et al*[23] deduced $\Theta_\omega = 680\ K$.

In Fig. 10,b we used a different approach[80,81,86], which is the data fit to Eq. 4, where $N$ is a free-fitting parameter (this approach was applied earlier[80] for compressed sulfur at pressures $P =$ $76, 86, 93\ GPa$). Deduced $\Theta_\omega(153\ GPa) = 743 \pm\ K$ and $N = 2.60 \pm 0.02$ are close, but not equal to $\Theta_\omega(153\ GPa) = 680\ K$ derived at the fixed $N = 3.0$ value (as reported by Osmond *et al*[23]). Based on this, there is an interest in revealing the $\Theta_\omega(P)$ and $N(P)$ values for all available $R(T, B = 0, P)$ data for compressed sulfur, which we showed in Figs. 10-12.

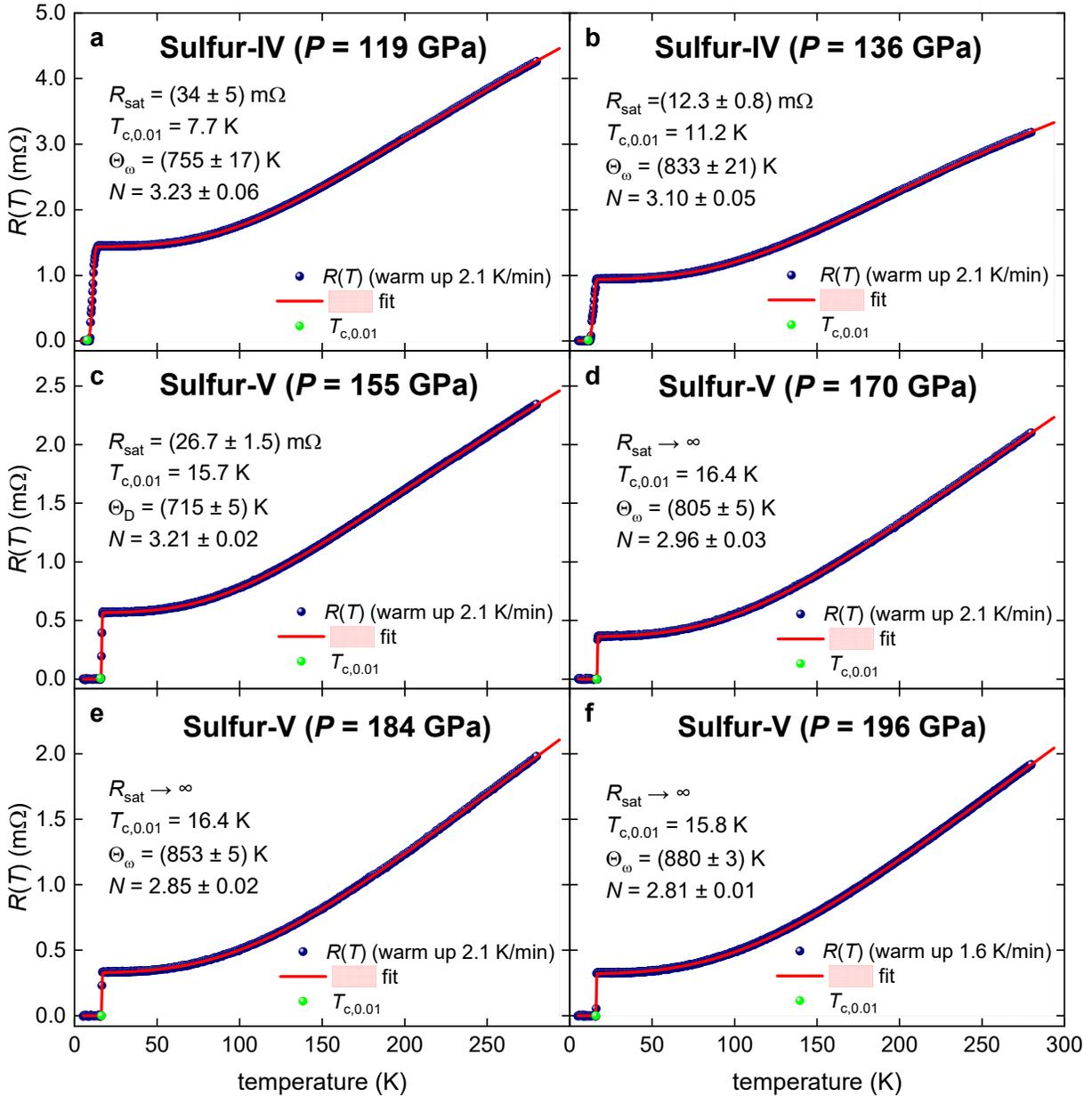

**Figure 11.** $R(T, B = 0, P)$ of the compressed sulfur measured by Wang *et al*[13] and data fits to Eq. 4, where $N$ is free-fitting parameter. Green balls indicate $T_{c,0.01}$. The goodness of fit for all panels is better than 0.9998. The thickness of 95% confidence bands (pink shadow areas) is narrower than the width of the fitting lines.

In our earlier study[80], we already fitted several $R(T,P)$ datasets reported by Shimizu et al.[2,5,61] to Eq. 4, where $N$ is free-fitting parameter, and in Figure 12 we show the fit for the remaining dataset reported in Refs.[2,5,61].

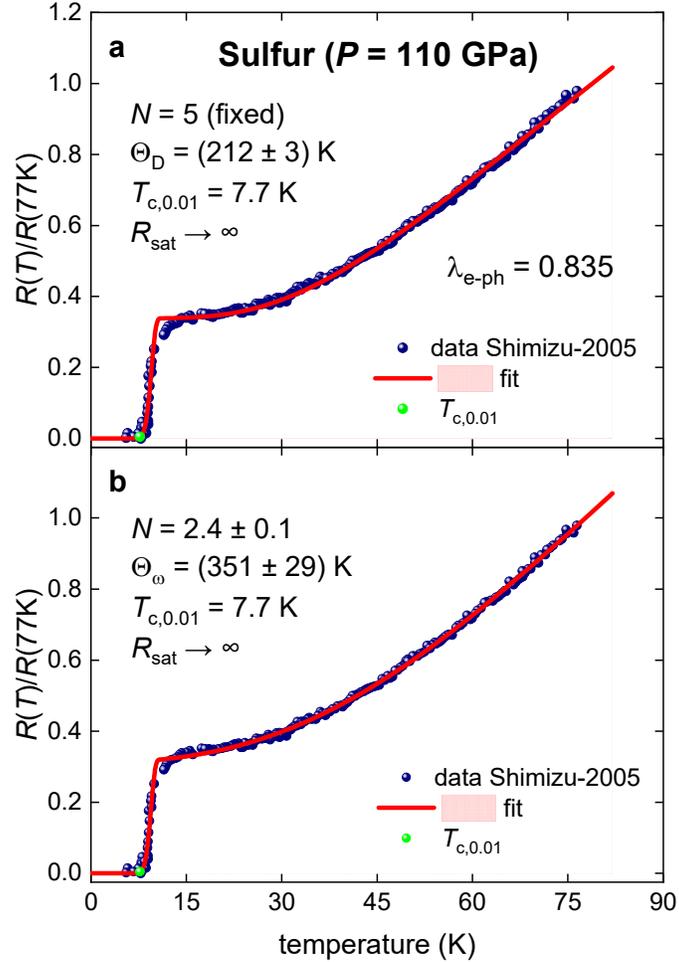

**Figure 12.** $R(T, B = 0, P = 110\ GPa)$ for compressed sulfur measured by Shimizu et al[5] and fits to Eq. 4, where $N$ is (a) fixed value ($N = 5$), and (b) is free-fitting parameter. Green balls indicate $T_{c,0.01}$. The goodness of fit (COD) is (a) 0.9972, and (b) 0.9978. The thickness of 95% confidence bands (pink shadow areas) is narrower than the width of the fitting lines.

In Figure 13 we summarized derived $T_c(P)$, $N(P)$, and $\Theta_\omega(P)$ dependences for highly compressed sulfur. Free-fitting power-law exponent is within a range of $2.4 \leq N \leq 3.2$, which is close, but not equal to the $N = 3$.

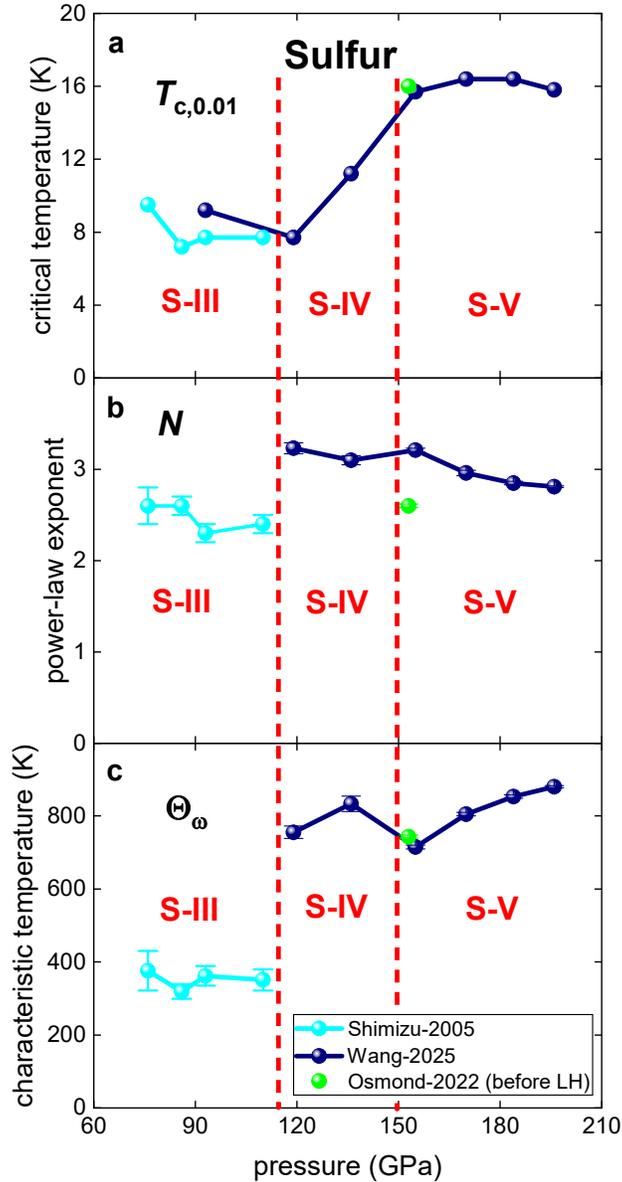

**Figure 13.** Derived (a) $T_{c,0.01}(P)$, (b) $N(P)$, and (c) $\Theta_\omega(P)$ for highly compressed sulfur. cyan – values deduced from experimental $R(T,P)$ data reported by Shimizu *et al.*[5]; navy – values deduced from experimental $R(T,P)$ data reported by Wang *et al.*[13]; green - value deduced herein from experimental $R(T, P = 153\ GPa)$ data reported by Osmond *et al.*[13,23]. Phase boundaries are based on Refs.[5,6,8–10,13,16,23].

### 3.6. Highly compressed unreacted sulfur in the H₃S+S mixture after laser heat

Osmond *et al*[23] also reported data for sample contained H₃S and S after the laser heating. Here is the instrumental broadening function $\beta_i(2\theta)$ in Figure 14, that was determined using X-ray scanning of a CeO2 sample. XRD peaks were fitted to Gaussian peak function and obtained $\beta_i(2\theta)$ dataset was fitted to Caglioti equation[48]:

$$\beta_{inst}(2\theta) = \sqrt{U \cdot tan^2\left(\frac{2\theta}{2}\right) + V \cdot tan\left(\frac{2\theta}{2}\right) + W}, \qquad (10)$$

where $U$, $V$, and $W$ are free-fitting parameters. Result of the fit is shown in **Error! Reference source not found.**4, where derived parameters are: $U = 0$ (fixed), $V = 0.0040 \pm 0.0004$, $W = 0.00284 \pm 0.00005$.

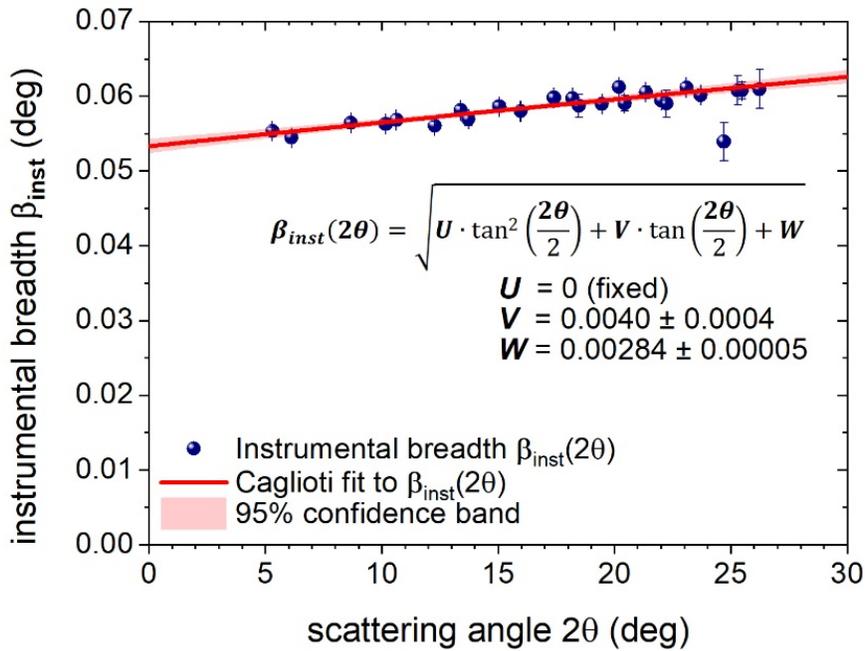

**Figure 14.** XRD peaks breadth, $\beta_{inst}(2\theta)$, for CeO2 recorded by Osmond et al[23] and the data fit to Caglioti equation (Eq. 10). Error bars are from the Gaussian fitting error. 95% confidence bands are shown by pink areas. Deduced parameters are shown. Fit quality is 0.8418.

In Figure 15, the blue circles indicate the points selected to analyze the strain distribution in sulfur, the original figure from Osmond et al[23] is designated there as Fig. 1,c.

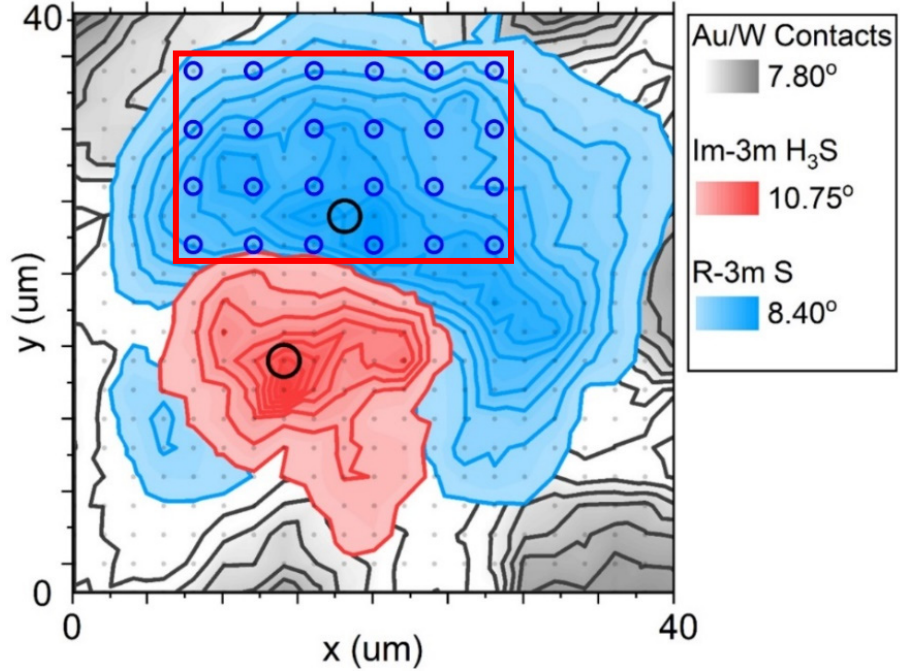

**Figure 15.** Selected points, depicted by blue circles (circles were shown in the original Figure 1,c from Osmond *et al*[23]), for which the size-strain analysis has been performed (XRD data for analysis was reported by Osmond *et al*[23]). The original figure from Osmond *et al*[23] is designated there as Fig. 1,c.

XRD peaks of sulfur were fitted to Gaussian peak function, and after the subtraction of the instrumental broadening, the obtained $\beta(\theta)$ dataset was fitted to Williamson-Hall equation[99]:

$$\beta(\theta) = \frac{k_s \times \lambda}{D \cos(\theta)} + 4 \times \varepsilon \times tan(\theta), \tag{11}$$

where $k_s$ is the Scherrer constant usually assigned as 0.9, $\lambda = 0.2894$ Å is the wavelength of the X-ray radiation used in Ref.[23], and $D$ is the mean size of nanocrystallines, and the $\varepsilon$ is the nanocrystalline strain.

Some of these fits are shown in Fig. 16. In the result, we found that the sulfur in the mixture of $H_3S+S$ in the DAC after laser heating has larger crystal size, $D$, which exceeds the resolution of the diffractometer (i.e. $D > 25\ nm$) and, thus, we revealed the sulfur strain map.

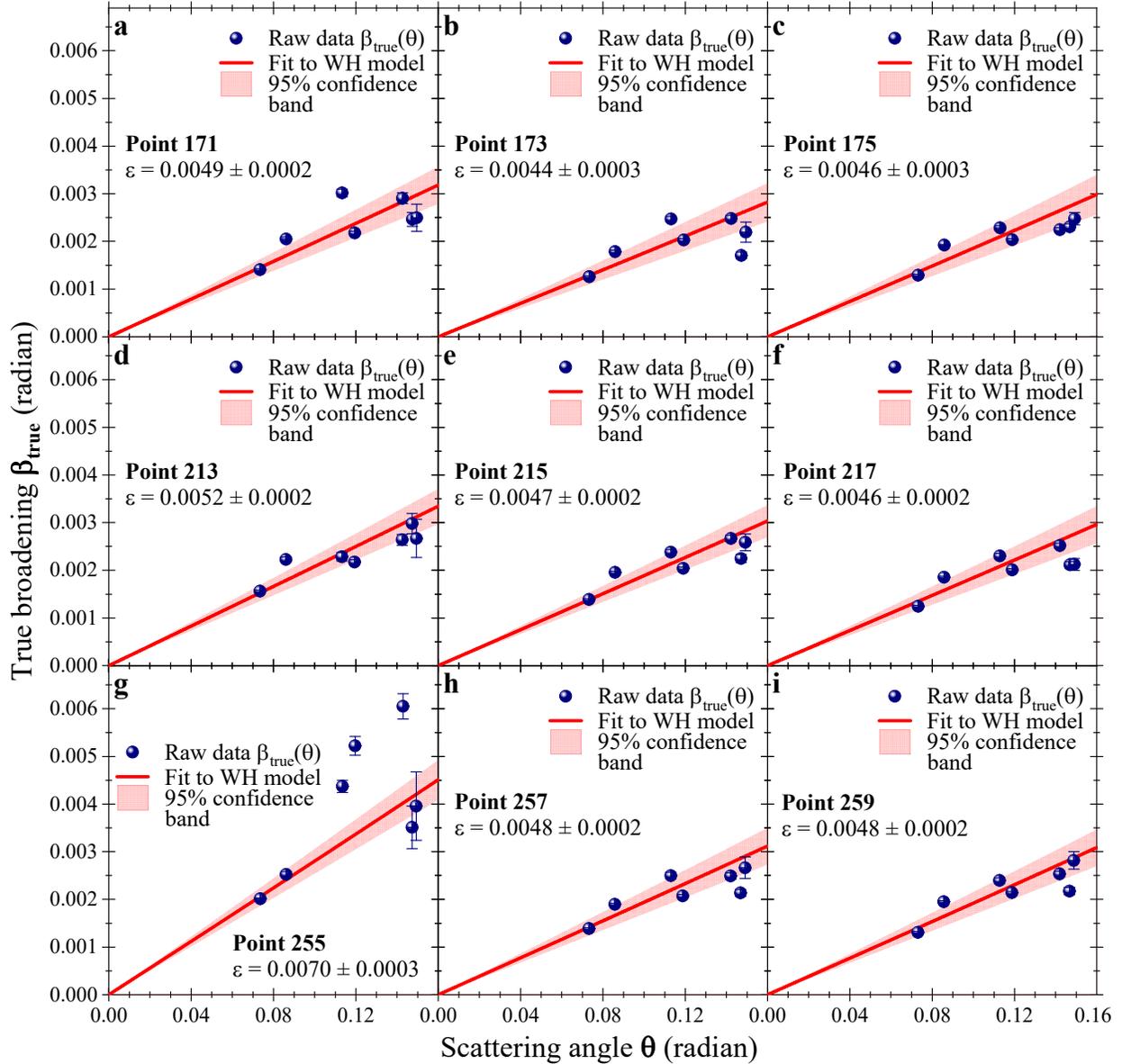

**Figure 16.** Several Williamson-Hall plots for mapping of strain ε(x,y) in sulfur region, marked by red frame in Fig. 15. Point designation is as it uses by Osmond *et al*[23] in their raw data files.

Figure 16 shows spatial maps of 8.4° sulfur XRD peak intensity and strain ε(x,y) in sulfur region, marked by red frame in Fig. 15, where X and Y coordinates correspond to both in Fig. 15. The strain is ranges from ε = 0.0045 to 0.0070 across analyzed area, where in the central region the strain is within a range of ε = (0.40-0.50) %. The comparing the two maps (showed in two panels of Fig. 17) it can be seen that areas of low sulfur peak intensity correlate with higher microstrain values. At the same time, the obtained values of microstrain in this experiment remain relatively low.

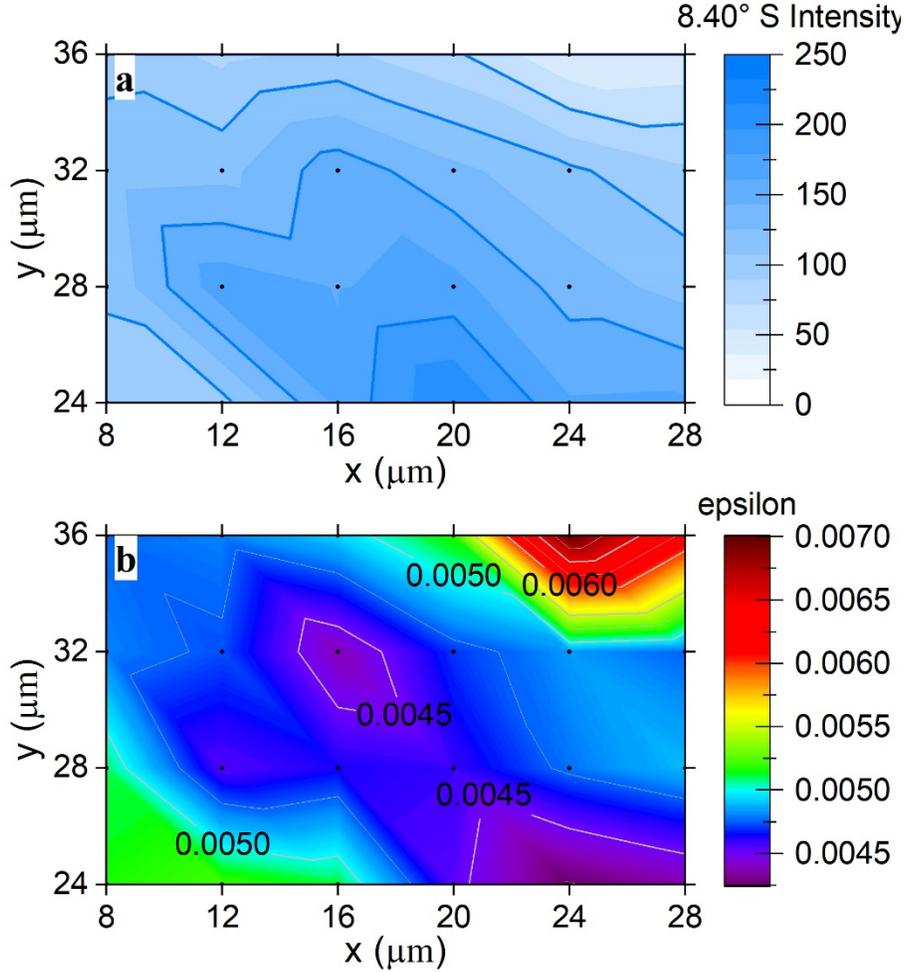

**Figure 17.** Spatial mapping of strain $\varepsilon(x,y)$ in sulfur region, marked by red frame in Fig. 15. X and Y coordinates correspond to both in Fig. 15: (a) intensity map of 8.4° sulfur XRD peak; (b) the $\varepsilon(x,y)$ map for sulfur extracted from XRD data reported by Osmond et al[23].

Osmond et al[23] also reported two $R(T, P = 153\ GPa)$ curves measured on samples with synthesized H$_3$S and remaining sulfur (after the laser heating of the mixture of the BH$_3$NH$_3$ and sulfur). In Figure 18, we showed the reported $R(T, P = 153\ GPa)$ datasets together with the fits to serial two-phase model which we also proposed herein:

$$R_{total}(T,P) = R_{sulfur}(T,P) + R_{H3S}(T,P) \qquad (12)$$

where each $R_{phase}(T,P)$ is described by Eq. 4 with independent parameters (i.e., $T_c^{onset}$, $\Theta_D$, etc.), and where we restricted the model to Bloch-Grüneisen[73,74] approach with $N = 5$.

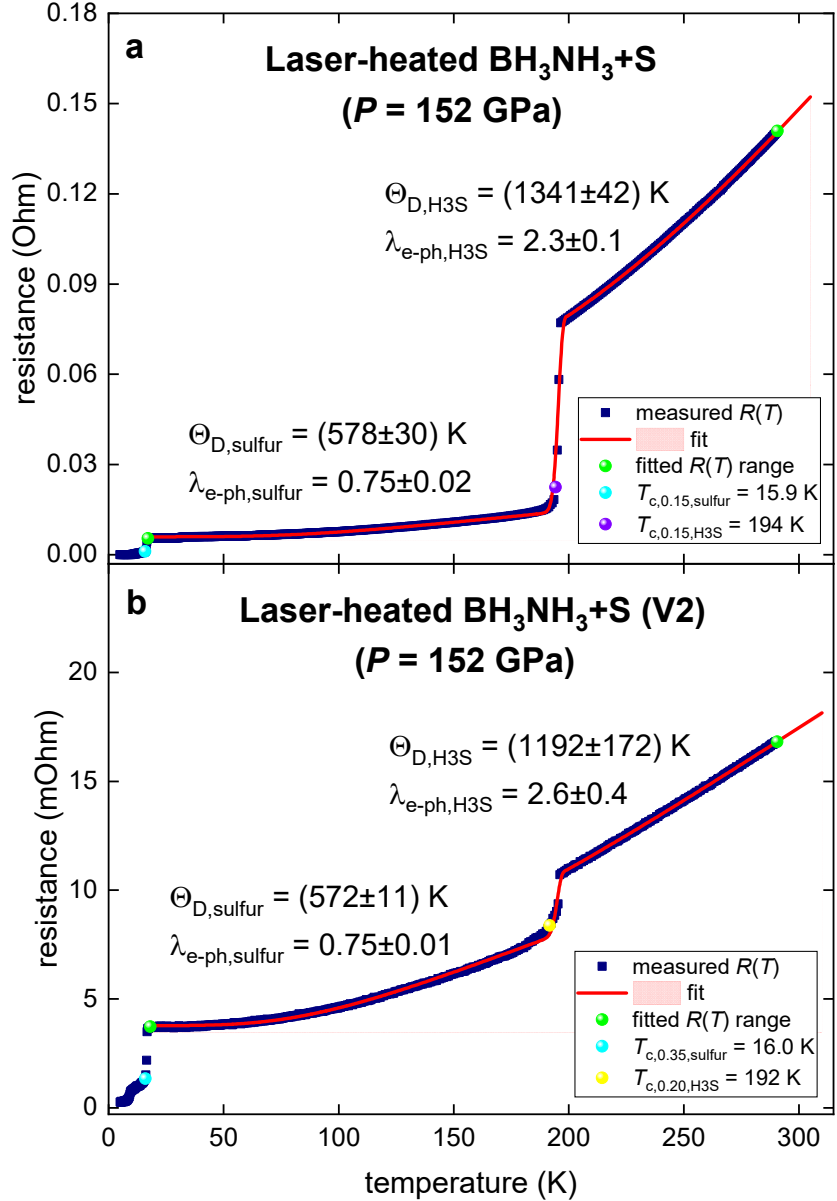

**Figure 18.** $R(T, P = 152\ GPa)$ for two samples of H$_3$S+S after laser heat (raw data reported by Osmond et al[23]) and fits to Eq. 12, where $N = 5$. Deduced parameters are shown in each panel. Green balls in both panels indicate $R(T,P)$ range used for fit. (a) Cyan ball indicates $T_{c,0.15}$ for sulfur; violet ball indicates $T_{c,0.15}$ for H$_3$S. (b) Cyan ball indicates $T_{c,0.35}$ for sulfur; violet ball indicates $T_{c,0.35}$ for H$_3$S. The goodness of fit (COD) is (a) 0.9996, and (b) 0.9985. The thickness of 95% confidence bands (pink shadow areas) is narrower than the width of the fitting lines.

It should be stressed, that $\Theta_{D,sulfur}(P = 152\ GPa)$ and $\lambda_{e-ph,sulfur}(P = 152\ GPa)$ deduced for the mixture of H₃S+S phases (after the laser heat of BH₃NH₃ and S precursors) are in excellent agreement with the respectful values derived for pure elemental sulfur (Fig. 6).

Determined $\Theta_{D,H3S}(P = 152\ GPa)$ and $\lambda_{e-ph,H3S}(P = 152\ GPa)$ are also in excellent agreement with values extracted from the $R(T,P)$ measured in sample with pure H₃S phase[68,83], as well as computed by the first-principles calculations[29,30,35,43,100].

### 3.7. Size-strain relation in fast ramp compressed polymeric sulfur

For the completeness of the overview of current status of high-pressure studies of sulfur, in Figure 19 we showed the Williamson-Hall plot for polymeric sulfur synthesized by fast ramp pressure technique to $P = 4.1\ GPa$ (XRD scan reported by Shi et al.[14] (XRD scan approximation by Gaussian peak functions is shown in Fig. 18,a).

Deduced size $D = (7.5 \pm 0.5\ \text{Å})$ and strain $\varepsilon = (0.03 \pm 0.01)$ should be associated with the cross-section of sulfur chains bundle. Considering that S-S distance in all sulfur phases is about[101] ~2.1 Å, and assuming that angle between sulfur atoms in the chains is ~120 degree, we can estimate that each bundle consists of $\frac{\pi}{4} \times \left(\frac{7.5\ \text{Å}}{2.1\ \text{Å} \times sin(30°)}\right)^2 \cong 40\ chains$. These chains are in strong distorted form, which can be proved by high level of strain, $\varepsilon = (3 \pm 1)\%$.

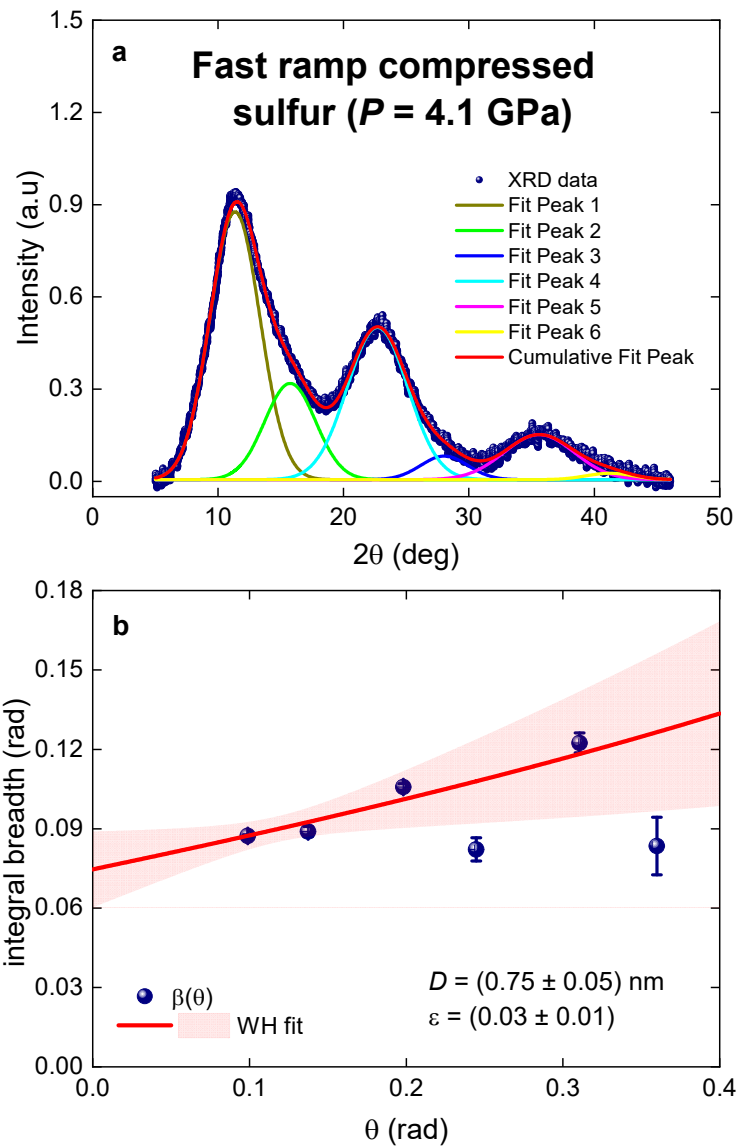

**Figure 19.** (**a**) XRD scan (data reported by Shi *et al.*[14]) for fast ramp compressed sulfur to $P = 4.1\ GPa$ and data approximation by Gaussian peak function. (**b**) Williamson-Hall plot for data presented in panel (**a**). The goodness of fit (R-squared COD) for (**b**) is 0.743. 95% confidence bands are shown by pink shadow area.

## 4. Conclusions

In this study we analyzed recently reported data on compressed sulfur[12–14,23]. The interest to this nonmetallic element, which has very versatile phase diagram[52,101], and which is converting to superconductor at high pressure, is remaining over the last five decades. Our main findings are:

1. Sulfur at $P = 160\ GPa$ has the relative jump of the specific heat at transition temperature $\frac{\Delta C_{el}}{\gamma T_c} = 1.8 \pm 0.1$, which is in a good agreement with reported[12] experimental value for the ratio $\frac{2\Delta(0)}{k_B T_c} = 3.89 \pm 0.06$.

2. Highly compressed sulfur in its elemental form and being compressed and laser heated in the mixture of H₃S and S, has practically identical Debye temperature, $\Theta_D(P)$, and the electron-phonon coupling constant $\lambda_{e-ph}(P)$.

3. Highly compressed sulfur-III and sulfur-V phases are in the intermediate band between conventional and unconventional superconductors in the Uemura plot. These phases have intermediate strength of the nonadiabaticity, i.e. $0.046 \leq \frac{\Theta_D}{T_F} \leq 0.2$, which is similar to MgB₂, pnictides, cuprates, La₄H₂₃, ThH₉, H₃S, LaBeH₈, and LaH₁₀.

4. Compressed superconducting sulfur in the mixture of of H₃S and S has low level of microstrain $\varepsilon \cong 0.5\%$, and high level of microstrain $\varepsilon \cong 3\%$ in the form of molecular chains fabricated by fast ramp compression.

**Acknowledgements**

The work was carried out within the framework of the state assignment of the Ministry of Science and Higher Education of the Russian Federation for the IMP UB RAS. E.F.T. gratefully acknowledge the research funding from the Ministry of Science and Higher Education of the Russian Federation under Ural Federal University Program of Development within the Priority-2030 Program.

**Conflict of interest**

The authors declared that they do not have any conflict of interest.

**Author contributions**

E.F.T. conceived the work. E.G.V.-Z. performed XRD data analysis. E.F.T. performed data analysis for resistance, upper critical field, and superconducting gap. E.F.T. wrote the manuscript with contributions from E.G.V.-Z.